\documentclass[prl, twocolumn, floatfix]{revtex4}
\usepackage{graphicx}
\usepackage{enumerate}
\usepackage{hyperref}
\usepackage{textcomp}
\usepackage{amsmath}
\usepackage{amssymb}
\usepackage{pgf}
\usepackage{booktabs}

\newcommand{\spmine}{1.10}
\newcommand{\myspace}{\edef\baselinestretch{\spmine}\Large\normalsize}

\begin{document}
\myspace

\title{Friction Anomalies at First-Order Transition Spinodals: 1T-TaS$_2$}

\author{Emanuele Panizon$^1$}
\author{Torben Marx$^4$} 
\author{Dirk Dietzel$^4$} 
\author{Franco Pellegrini$^{1}$} 
\author{Giuseppe E. Santoro$^{1,2,3}$}
\author{Andre Schirmeisen$^4$} \email[E-mail: 
]{schirmeisen@ap.physik.uni-giessen.de}
\author{Erio  Tosatti$^{1,2,3}$} \email[E-mail: ]{tosatti@sissa.it}

\affiliation{$^1$International School for Advanced Studies (SISSA), Via Bonomea
  265, 34136 Trieste, Italy}
\affiliation{$^2$CNR-IOM Democritos National Laboratory, Via Bonomea 265, 34136 
Trieste, Italy}
\affiliation{$^3$The Abdus Salam International Centre for Theoretical Physics 
(ICTP), Strada Costiera 11, 34151 Trieste, Italy}
\affiliation{$^3$Institute of Applied Physics, Justus-Liebig-University Giessen, 
Heinrich-Buff-Ring 16, 35392 Giessen, Germany}

\begin{abstract}
\textbf{Revealing phase transitions of solids through mechanical anomalies in 
the friction of nanotips sliding on their surfaces is an unconventional and 
instructive tool for continuous transitions, unexplored for first-order ones. 
Owing to slow nucleation,  first-order structural transformations generally do 
not occur at the precise crossing of free energies, but hysteretically, near the 
spinodal temperatures where, below and above the thermodynamic transition 
temperature, one or the other metastable free energy branches terminates. The 
spinodal transformation, a collective one-shot event with no heat capacity 
anomaly, is easy to trigger by a weak external perturbations. Here we propose 
that even the gossamer mechanical action of an  AFM tip may locally act as a 
surface trigger, narrowly preempting the spontaneous spinodal transformation, 
and making it observable as a nanofrictional anomaly. Confirming this 
expectation, the CCDW-NCCDW first-order transition of the important layer 
compound 1T-TaS$_2$ is shown to provide a demonstration of this effect.}

\end{abstract}

\maketitle

\section{Introduction}

The development of fresh theoretical and experimental tools aimed at revealing 
and understanding solid state phase transitions through their surface 
nanomechanical and nanofrictional  effects is an ongoing unconventional, yet 
very useful approach.   Friction of nanosized tips on dry solid surfaces has 
been proposed to represent what one might term ``Braille spectroscopy'' -- 
reading the physics underneath by touching \cite{vanossi2013colloquium}. 
For second-order, continuous phase transitions, a notable example has been the 
detection of displacive structural transformations as reflected by AFM 
dissipation anomalies caused by critical fluctuations, predicted 
\cite{benassi2011sliding} and observed in noncontact friction on SrTiO$_3$ 
\cite{kisiel2015noncontact}. Another non-structural example is the drop of 
electronic friction, observed upon cooling a metal below  the superconducting Tc 
in correspondence to the opening of the BCS gap \cite{kisiel2011suppression}. 
The injection of a 2$\pi$ phase slip in the local order parameter of an 
incommensurate phase is an additional interesting event that can be triggered by 
an AFM tip.\cite{langer2014giant}

A vast majority of solid state structural and electronic phase transitions is,  
however, of discontinuous, first-order type. Should one expect a frictional 
anomaly at the surface of a solid which undergoes a first-order structural 
transition? Lacking critical fluctuations, that frictional signature might it 
not just consist of some unpredictable and unremarkable jump? This scenario is, 
we propose, unduly pessimistic, countering that not one but two frictional 
anomalies are to be expected at a first order transition. They should occur at 
the hysteresis end-point temperatures, where both heating and cooling 
transformations are close in character to spinodal  -- the point where the 
dissolution of a metastable state takes place. At these two temperatures, on 
both sides of the thermodynamic transition temperature, an Atomic Force 
Microscope/Friction Force Microscope (AFM/FFM) dissipation peak is to be 
expected as the tip moves on, sweeping  in the course of time newer and newer 
surface areas where the near-spinodal transformation can be ``harvested''. These 
predictions are first argued theoretically and then demonstrated experimentally 
in the important layer compound 1T-TaS$_2$.   

\section{Mean-field theoretical model}

Beginning with theory, we adopt the simplest mean-field Landau-Ginzburg-Wilson 
\cite{langer1967theory} or Cahn-Allen \cite{cahn1977microscopic} approach, which 
works reasonably well for many structural transitions. Assume the schematic 
model bulk solid free energy density 

\begin{equation}
f[\Psi] = -\frac{r}{2} \Psi^2(\rho) + \frac{u}{4}\Psi^4(\rho) + 
\frac{J}{2}(\nabla \Psi)^2 + h(\rho)\Psi(\rho),
\end{equation}
(where $r$, $u$, $J$ are positive parameters) governing the evolution of a 
generic, non-conserved real order parameter  $\Psi$ supposed to represent 
collectively all mechanically relevant thermodynamic variables, as  a function 
of spatial coordinate $\rho$ (in this schematic outline, we provisionally ignore 
the distinction between surface and bulk). The external field $h$ includes here 
a  uniform term describing the free energy imbalance between the two minima at 
negative and positive $\Psi$  ($h$ thus represents here the temperature 
deviation from the  first-order thermodynamic transition point)  plus a 
localized mechanical perturbation representing the tip which, when moving in the 
course of time, will undergo mechanical dissipation, observable as friction. In 
the spatially uniform, field-free case $(\nabla \Psi)^2=0$, $h=0$,  two 
equivalent free energy minima $F_0 = - r^2/4u$ occur at 
$ \Psi_0^{\pm} = \pm \sqrt{r/u}$ identifying the two phases. A first order 
transition occurs between them when a growing  uniform $h$ causes $\Psi$ to 
switch from initially positive to negative or viceversa. For the transition to 
occur near $h=0$  nucleation is required, allowing thermal crossing of the large 
free energy barrier between the two nearly equivalent states. Nucleation is a 
generally slow process so that a $\Psi>0$ metastable state often persists up to 
large positive fields $h$. Upon reversing the field, the $\Psi<0$ state can 
similarly persist for negative $h$, giving rise to hysteresis. The maximum 
theoretical width of the hysteresis cycle is determined by the two 
\textit{spinodal} points $h_s = \pm 2r^{3/2}/3^{3/2}u^{1/2}$,  where the 
transformation must necessarily take place because at each of them the 
metastable free energy minimumum disappears, as sketched in Fig.1(a). At the 
spinodal points  $\Psi_s = \pm \sqrt{r/3u}$ and $f_s = (1/12) r^2/u$ the 
transformation occurs collectively rather than locally, as amply described in 
literature, reviewed in a different context in  \cite{binder1987theory}. 

\begin{figure}[!htp]
\centering
\includegraphics*[width=0.45\textwidth,angle=0]{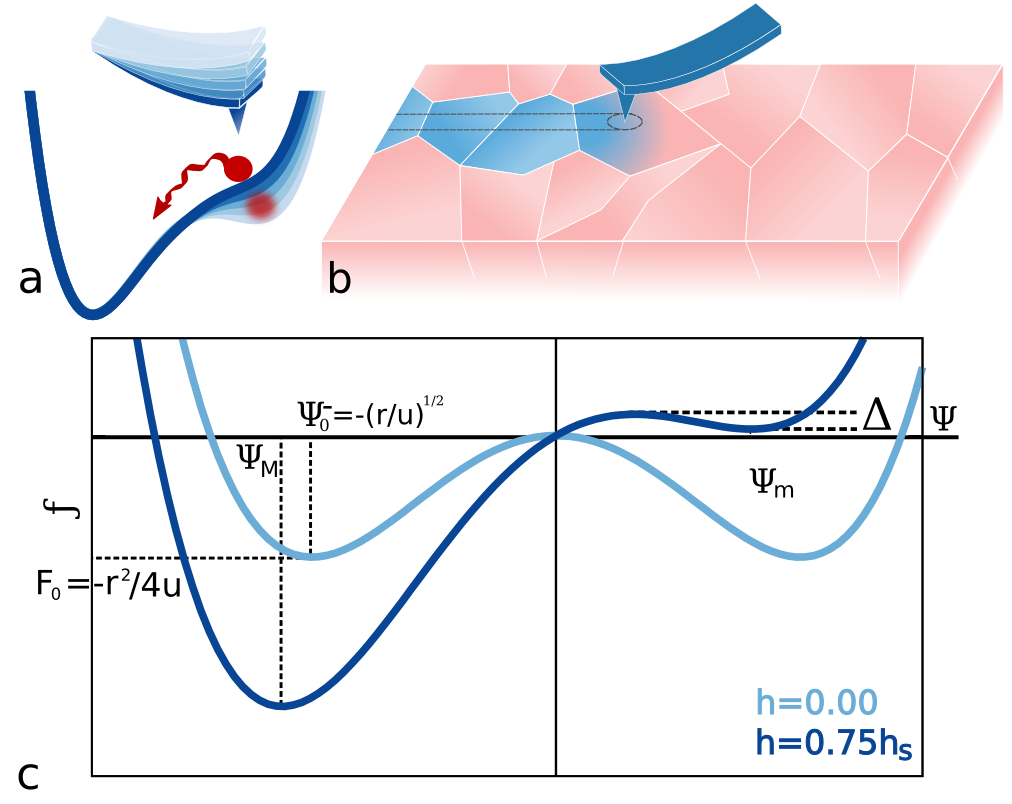}
\caption{a) Schematic representation of free energy behaviour versus order 
parameter near a spinodal point, and triggering action of an external tip; b)  
schematic of a sweeping tip on a solid with domains, triggering the local 
metastable-stable transformation and leaving behind a transformed trail; c) 
parameters of the free energy density near a spinodal point.}
\label{fig:sketches}
\end{figure}

A spinodal point is associated with the collective dynamics of all macroscopic 
variables accompanying the first order transition including structure, volume, 
conductivity, etc. As that point is approached, even a small perturbation can 
locally overcome the marginal free energy barrier and  trigger a large-scale 
transformation from the metastable to the stable state, as pictured in Fig. 
\ref{fig:sketches}(a). If that perturbation is provided by a sliding tip, the 
small but finite triggering work  will show up as a a frictional dissipation 
burst.  As the tip moves on, it can convert newer and newer patches from 
metastable to stable, Fig. \ref{fig:sketches} (b). The pursuit of the frictional 
consequences of a first order transition close to its spinodal points is our 
goal.

Starting with $\Psi>0$, and turning up the uniform field (i.e., lowering the 
temperature approaching from above the real system spinodal temperature on 
cooling down) $h\rightarrow h_s$, we observe that $f[\Psi]$ is a local minimum 
--  a metastable state -- protected by a marginal barrier $\Delta$ which 
disappears at the spinodal point $h = h_s$, Fig. \ref{fig:sketches}(c). Before 
that point is reached, in the metastable state, a weak static local perturbation 
$\delta h(\mathbf{\rho})$ imparts to the state variable $\Psi$ a local 
modification, whose effect, initially small, grows as the spinodal point is 
approached. If moreover the perturbing agent, in our case the nano-tip, moves in 
space with  velocity $\mathbf{v}$, so that 
$h_{\rm tip}(\mathbf{\rho},t)=h_0(\mathbf{\rho}-\mathbf{v}t)$, then it may or it may 
not succeed to locally trigger the spinodal transformation. If it does, then 
some mechanical work will be spent, and that expense will reflect in the form of 
a  burst in the tip's mechanical dissipation. 

\begin{figure}[!htp]
\centering
\includegraphics*[width=0.4\textwidth,angle=0]{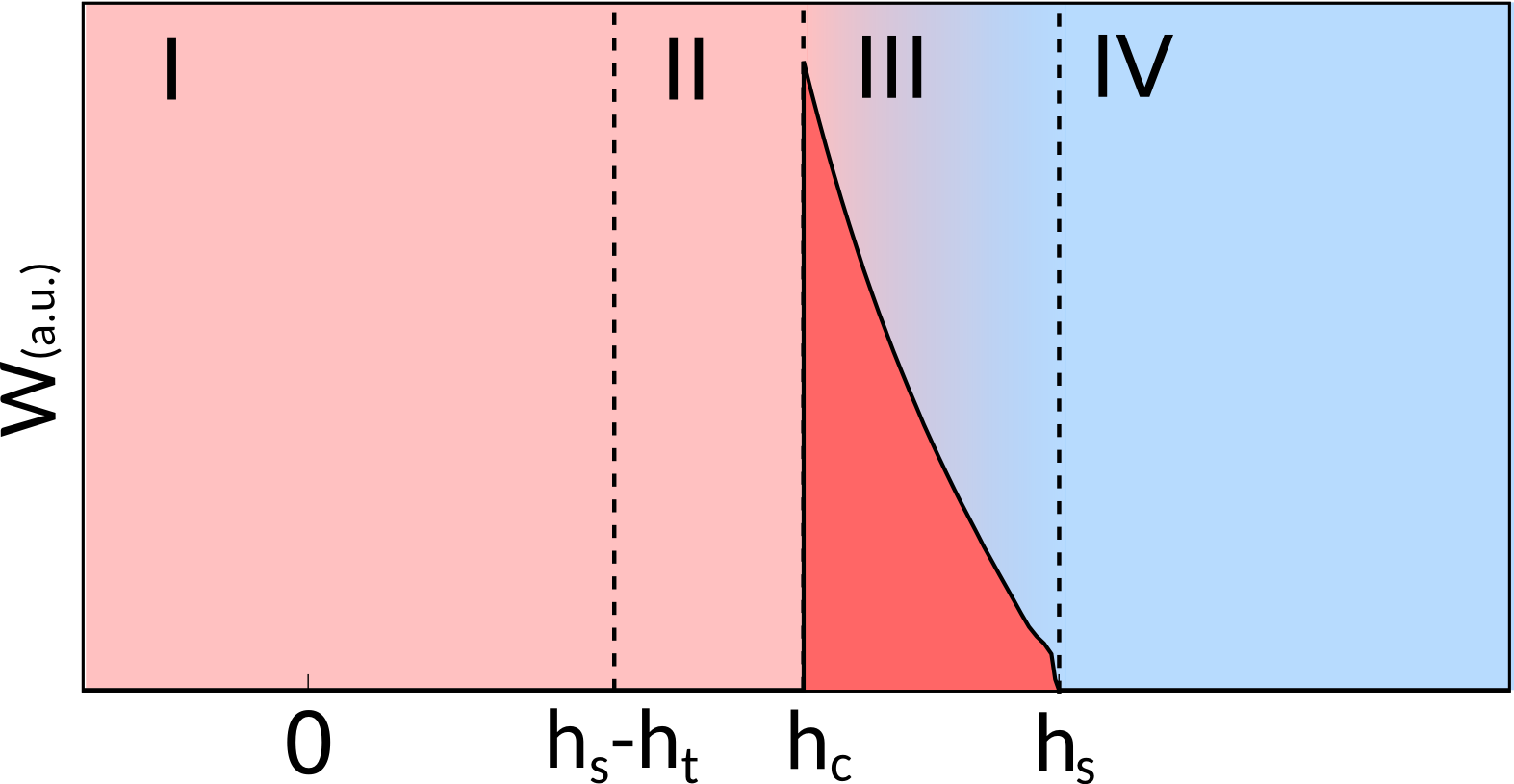}
\caption{Predicted tip dissipation $W$ as a function of $h$ (red area). In the 
cooling example, increasing $h$ stands for decreasing temperature, and $h =h_s$ 
is the spinodal point. The sharp dissipation peak occurs at the threshold 
temperature where the tip perturbation succeeds in triggering the metastable 
state's demise, thus preempting the spontaneous spinodal transformation before  
$h_s$ is reached. The threshold $h_c$ depends on details including the tip 
nature, radius, and load.  In this figure the tip radius was $R_{\rm tip} \sim 1.1 
\sqrt{r/u}$, other parameters were $u=10$, $r=10$ and $j=1$. The regions I, II, 
III and IV are defined in text.} 
\label{fig:dissipation}
\end{figure}

Four different frictional regimes are crossed as a function of $h$ (i.e., of 
temperature) -- for example when evolving from a  high temperature metastable 
state $\Psi_m$ to a low temperature stable state $\Psi_M$ on cooling. In regime 
(I), $h$ is still far from the spinodal point $h_s$, the free energy barrier 
protecting the metastable state is substantial, the tip perturbation is too weak 
to push the system over it, and the tip friction is unaffected. In a second 
regime (II), the tip may succeed to ``wet'' its surrounding with a small 
converted nucleus $\Psi_M$ of radius  $R_{\rm tip}$, yet still unable to overcome 
the nucleation barrier if $R_{\rm tip}<R_c$, the effective ``inhomogeneous''  
nucleation critical radius. Depending whether this nucleus does or does not 
reconvert back to $\Psi_m$ as the tip moves on, there will or will not be 
frictional work. Assuming reconversion (for slow tip motion), the friction is 
again zero, because the transformed nucleus is carried along adiabatically by 
the tip. In regime (III), as  $h$ increases, the nucleation radius eventually 
gets smaller than the tip perturbation radius, $R_c < R_{\rm tip}$ . The system 
suddenly overcomes the barrier as in Fig. \ref{fig:sketches}(a), thus provoking 
the irreversible transformation $\Psi_m \to \Psi_M$ extending in principle out 
to infinite distance -- in practice, out to some macroscopically determined 
radius $L$ defined by the sample quality, defects, and morphology. At this 
threshold temperature the mean tip frictional dissipation will suddenly jump 
from zero to finite, thereafter decreasing smoothly and eventually vanishing 
when the true spinodal point $h \rightarrow h_s$ is reached, and dissipation 
again disappears, regime (IV). Our model predicts the frictional dissipation 
burst in the shape shown Fig. \ref{fig:dissipation}, a behaviour which we now 
describe in our model before seeking an experimental demonstration.

Consider a configuration where, as in classical nucleation theory (CNT) the 
system is forced to evolve from  $ \Psi =\Psi_M$ at $\rho$=0, to $ \Psi =\Psi_m$ 
at $\rho=\infty$. A trial function that shows this behaviour is constructed as 
$\Psi(\rho) = \Psi_m + (\Psi_M-\Psi_m)/2 \tanh ((\rho - R_0)/\gamma)$, where 
$R_0$ is the radius of the droplet and $\gamma$ its interface width. We consider 
the nucleation barrier $F[\rho;R_0,\gamma]$, which depends variationally on the 
two parameters $R_0$ and $\gamma$. Based on that we can numerically calculate 
the homogeneous nucleation radius $R_c$ and its corresponding barrier. Now add 
to $F[\rho]$ the tip perturbation $h_{\rm tip}({\rho}) = h_{\rm tip} \Theta( 
\Vert{\rho}-\textbf{x}_{\rm tip}\Vert-R_{\rm tip})$ whose effect is to lower the local 
barrier as sketched in Fig. \ref{fig:sketches}(a). At $h=h_c$ the nucleation 
radius becomes smaller than the wetting radius $R_c<R_{\rm tip}$, the local 
nucleation barrier disappears and the massive  transformation is triggered. The 
tip will spend at that point the one-shot triggering work  $ W = F_0 $. This 
work, as mentioned, is paid only once, because after conversion  the stable 
phase $\Psi_M$ extends macroscopically away, and the system becomes subsequently 
insensitive to the tip. It should be stressed here that, unlike second order 
transitions between equilibrium states, which take place reversibly as the 
temperature is cycled across the critical point, the spinodal transformation 
takes place only once as the spinodal point is first crossed (unless the system 
is, as it were, ``recharged'').  In a real system, however, the size of the 
transformed region is limited by defects to some average radius  $L$ determined 
by, e.g., grain boundaries, steps, etc., so that newer and newer metastable 
surface areas can be ``harvested'' in the course of time, as sketched in (Fig. 
\ref{fig:sketches}(b)). The  tip  moving with velocity $v$  will  explore fresh 
untransformed metastable regions with a rate $\mu \sim v/L$, (Fig 1(c)) 
therefore dissipating a frictional power  $P= W \mu =F_0 v/L $, a quantity which 
is nonzero in the temperature range corresponding to $h_c < h < h_s$ as depicted 
 in Fig. \ref{fig:dissipation}.

\section{Experimental verification in 1T-T\lowercase{a}S$_2$} 

\begin{figure}[!htp]
	\centering
	\includegraphics*[width=7cm,angle=0]{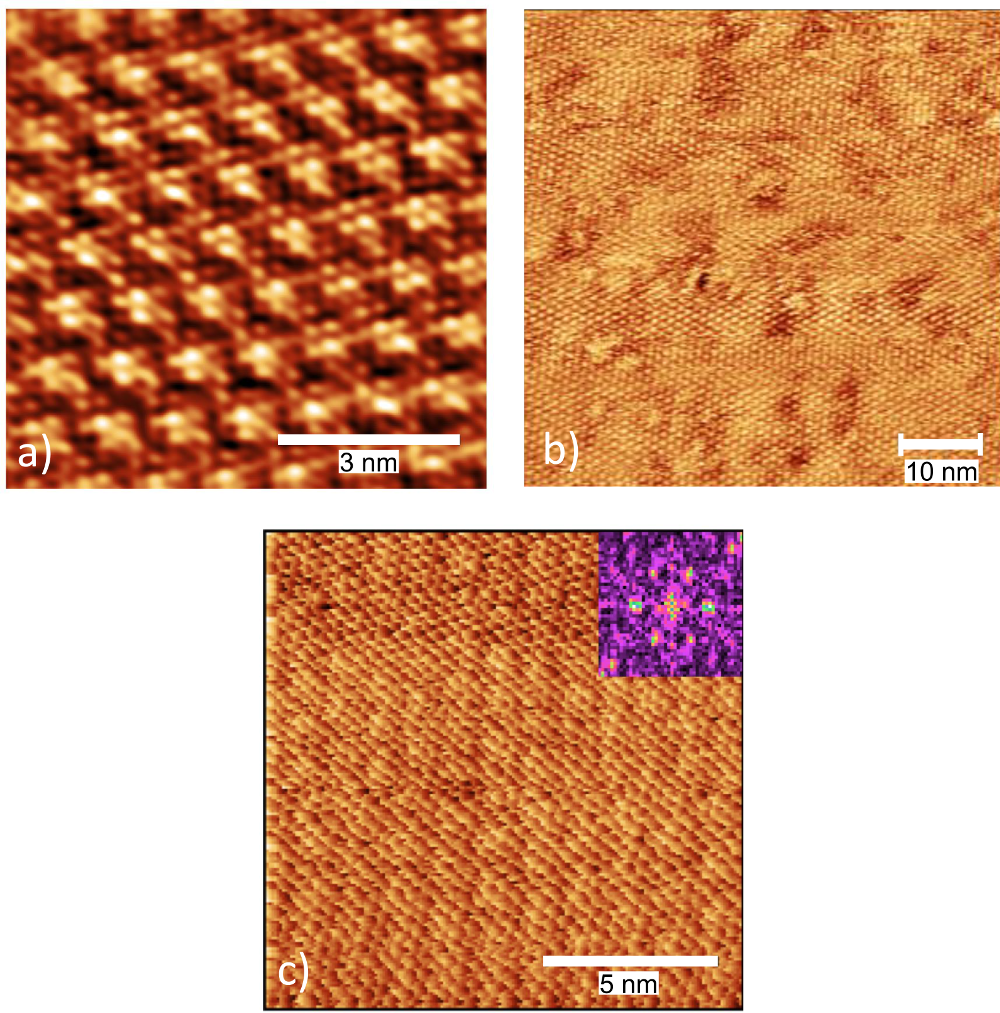}
	\caption{a) Atomically resolved STM image (8x8nm$^2$) of 1T-TaS$_2$ 
(296K, $I_T$=1nA, $U_T$=5mV). The $\sqrt{13}\times\sqrt{13}$  superstructure can 
clearly be identified and the experimental lattice constant of a=1.1nm agrees 
with literature \cite{Thomson1994STM}. b) STM image (60x60 nm$^2$ recorded in 
the CCDW phase at 161K ($U_T$=5mV, $I_T$=0.3nA). The STM image reveals typical 
sizes of defect free regions of a few nm$^2$. c) Friction force microscopy image 
obtained in the CCDW phase. A clear atomically resolved stick slip pattern is 
visible. The FFM image show only miniscule contrast for the superstructure, 
better visible in the Fourier transformation showing 6 bright spots (inset). 
Other frames, unlike this very perfect one,  show defects and imperfections   
which set the effective length scale $L$ discussed in text.}
	\label{fig:highres}
\end{figure}

Thus far the theory. To verify predictions in a well defined, physically 
interesting case we choose a thermally driven structural transition with an 
established  hysteresis cycle between phases that do not differ too strongly 
from one another. Such is the case of the transitions in the celebrated layer 
compound 1T-TaS$_2$ between a low-temperature Commensurate Charge-Density-Wave 
(CCDW) $\sqrt{13}\times\sqrt{13}$ phase, \cite{wilson1975charge, sipos2008mott} 
believed to be Mott insulating \cite{tosatti1976nature, fazekas1979el}, and a 
Nearly Commensurate Charge-Density-Wave (NCCDW) phase, metallic and even 
superconducting under pressure ~\cite{sipos2008mott}. 
This first-order transition takes place reproducibly in a single stage near 
$T_{NC} \sim$ $173$ K upon cooling, and in two stages, at  $T_{CT} \sim$ 223 K, 
$T_{TN} \sim$ 280 K, upon heating. That very reproducible hysteresis pattern, 
partly reproduced from Ref.~\cite{sipos2008mott}  in Fig.~\ref{fig:highres}, 
suggests  that $T_{NC}$ and $T_{CT}$ are, to a good approximation, spinodal points 
of 1T-TaS$_2$. That is strongly confirmed by very recent heat capacity data by 
Kratochvilova et al.~\cite{kratochvilova2017}, showing no anomaly at $T_{NC}$ and 
$T_{CT}$, where at the same time large electrical and structural bulk transformations 
take place. A spinodal transformation occurs, upon temperature cycling $\pm \Delta T$, 
only once, thus averaging all internal energy effects to zero upon repeated passage. 

It should also be mentioned that 1T-TaS$_2$
and its phases are and have been the subject of very intense studies over the last five years, 
in connection especially with transient or hidden metastable phases under high 
excitation \cite{Perfetti06,Vaskivskyi14, Laulhe17}  
and/or in connection with unusual substrate, thickness, and disorder dependence 
of its phase transitions \cite{Yoshida14,Yoshida15,Hovden16,Zhao17}. 
To begin with, we restrict here to bulk 1T-TaS$_2$ in equilibrium. Focusing for definiteness on 
the NCCDW $\leftrightarrow$ CCDW transition upon cooling, and consider the 
phenomena which we might expect  in AFM/FFM friction measurements as temperature 
crosses that transition. First, frictional heat dissipation into the substrate 
(phononic friction) could in principle differ in the two phases, because their 
structures, phonon spectra, mechanical compliances are, even if mildly, 
different -- for example, the NCCDW structure possesses a network of  
``soliton'' defects, absent in the CCDW. Second, electronic friction due to 
creation of electron-hole pairs could be present in the NCCDW phase which is 
metallic, and not in the CCDW which is insulating. Both mechanisms do suggest a 
higher noncontact friction in the NCCDW phase above $T_{NC} \sim$ $173$ K than 
in the CCDW phase below that.  Our experiment however measures hard contact 
friction, where these contribution turn out to be undetectable. The third, and 
central dissipation route described earlier is the main frictional feature which 
we observe near the spinodal points. 

In our friction force microscopy experiments we used 1T-TaS$_2$-flakes with a 
size of approximately $4\times4$ mm$^2$ and a thickness before cleaving of about 
50$\mu$m. To yield clean surface conditions, the samples were freshly cleaved 
directly before transfer to the UHV chamber of a commercial Omicron-VT-AFM/STM 
system. Inside the UHV chamber the samples were additionally heated to 
100$^\circ$C for 1h to remove residual adsorbates from the surface. To ensure 
that the sample is apt to detect the anticipated effects, we first used high 
resolution STM imaging to identify the most characteristic feature related to 
the NCCDW and CCDW, namely the $\sqrt{13}\times\sqrt{13}$ superstructure formed 
by 13 Ta-atoms arranged in a star shape around a central atom 
\cite{wilson1975charge}. This superstructure is revealed in Fig. 
\ref{fig:highres}a, measured at 296K. At this temperature, the superstructure 
forms separate hexagonal domains, which coalesce during the phase transformation 
on cooling, leading to  CCDW-1T-TaS$_2$ \cite{sipos2008mott}. 

Subsequently we use high resolution friction force microscopy (FFM) for a first 
analysis of the sample with respect to tribological properties. Atomically 
resolved stick-slip is regularly observed and Fig. \ref{fig:highres}b shows an 
example, which was measured in the CCDW phase at 173K using a standard 
Si-cantilever (Nanosensors LFMR, normal force constant k=0.2N/m). Discerning the 
superstructure from the lateral force data in Fig. \ref{fig:highres}b is 
difficult but can be achieved by calculating the Fourier transform, where the 
superstructure leads to characteristic bright spots as shown in the inset of 
Fig. \ref{fig:highres}b. 

\begin{figure}[!htp]
	\centering
	\includegraphics*[width=7.5cm,angle=0]{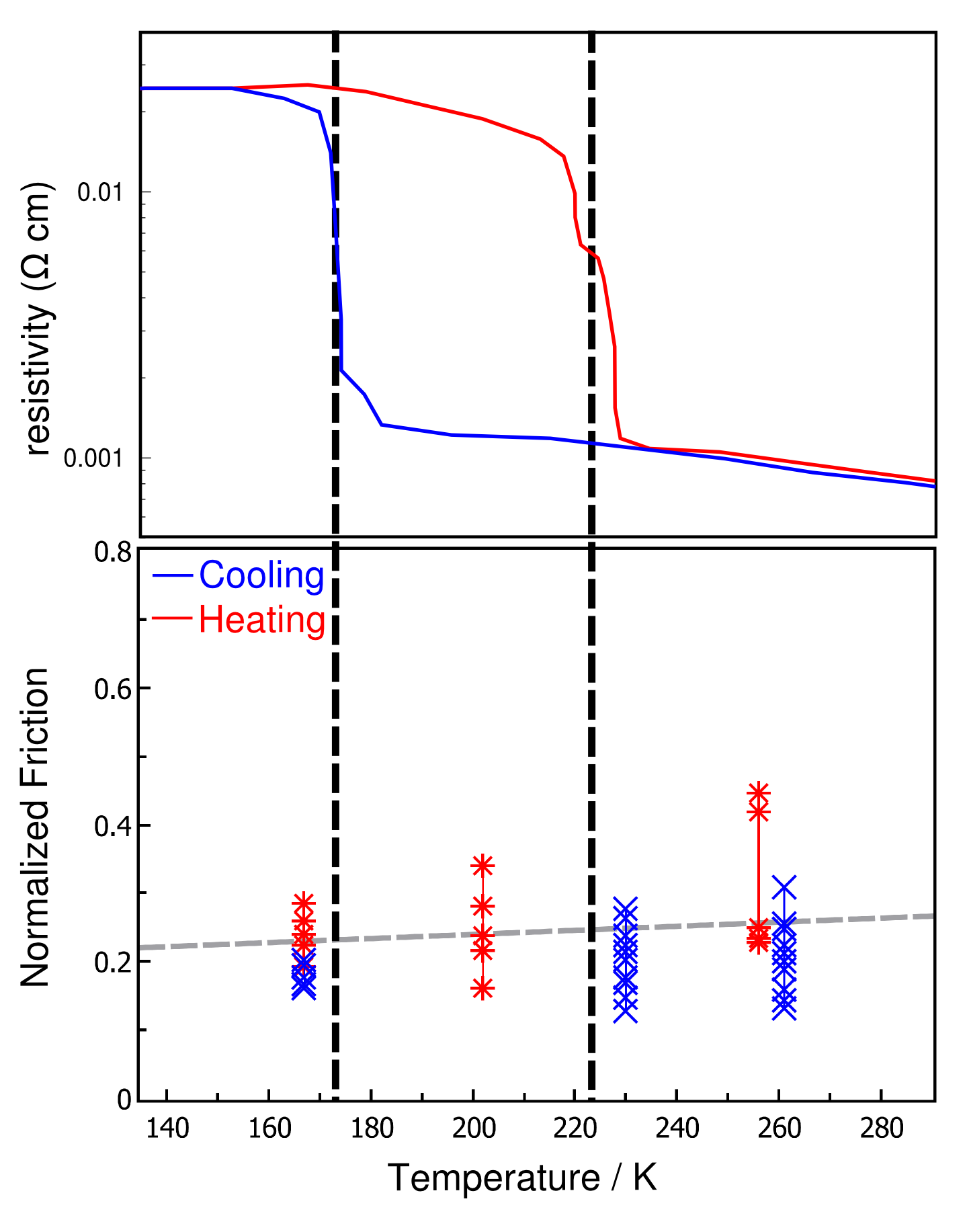}
	\caption{Top) Bulk resistivity of 1T-TaS$_2$ versus temperature across 
the CCDW to NCCDW transitions, reproduced from Ref. \cite{sipos2008mott}. Black 
dashed lined mark $T_{NC} \sim$ $173\pm2$K and $T_{CT} \sim 223$K on cooling 
and on heating respectively, transitions that are spinodal in character. 
Bottom) Coarse-scale temperature dependence of FFM friction relative to the 
average room temperature value,  measured during cooling and heating. 
 No direct correlation between friction and the change of electrical or of 
structurally commensurate or incommensurate characters is found  across $T_{NC}$ 
and $T_{CT}$ within experimental error.}
	\label{fig:TRange}
\end{figure}

In the first experimental run, which serves as a coarse-scale reference, we 
measure the temperature dependence of friction over a wide temperature range 
from room temperature down to 160K. Fig. \ref{fig:TRange}(bottom) shows that the 
friction remains constant within errors in the relevant range from 160K to 260K. 
In particular, there are no discontinuities  across the transition temperatures 
$T_{NC} \sim$ $173\pm2$ K and $T_{CT} \sim$ 223 K indicated by the dashed lines. 
Also there is no hysteresis between the cooling and heating cycles. One can see 
that incommensurability and metallization do not impact friction on this coarse 
scale. A much more detailed study is necessary to discern the influence of the 
hysteretic (spinodal) transformations, present in structure and in conductivity, 
on the tip friction. 

We therefore focus on the friction signal in a narrow temperature window around 
the anticipated spinodal transition points. We use a specific experimental 
protocol to measure lateral forces while crossing the transitions. First, the 
NCCDW to CCDW transformation is analyzed during cool down. For this the sample 
temperature was set to a constant value slightly above the transition point 
(appr. 195K). Once a stable sample temperature is established, continuous 
scanning of FFM images with a size of $50\times50$ nm$^2$ at a normal force 
set-point of $F_N=14$ nN and a scan speed of $v_{\rm scan}=250$ nm/s is started. 
Then the sample temperature is slowly reduced at a rate of appr. $0.2$ K/min 
until the minimum temperature of 170K is reached, while the scanning is 
continously running with the normal force feedback enabled. The temperature 
change induces a z-drift of the sample, and therefore only a small temperature 
window of about 10-20K is accessible with this method. Once the sample has been 
cooled down to the CCDW phase, the same procedure was used to analyse the 
transition from CCDW to NCCDW during heating. Here, 215K was chosen as a 
starting point and the temperature was increased at a similar rate up to 225K, 
thereby spanning the full phase transition. 

\begin{figure}[!htp]
	\centering
	\includegraphics*[width=6cm,angle=0]{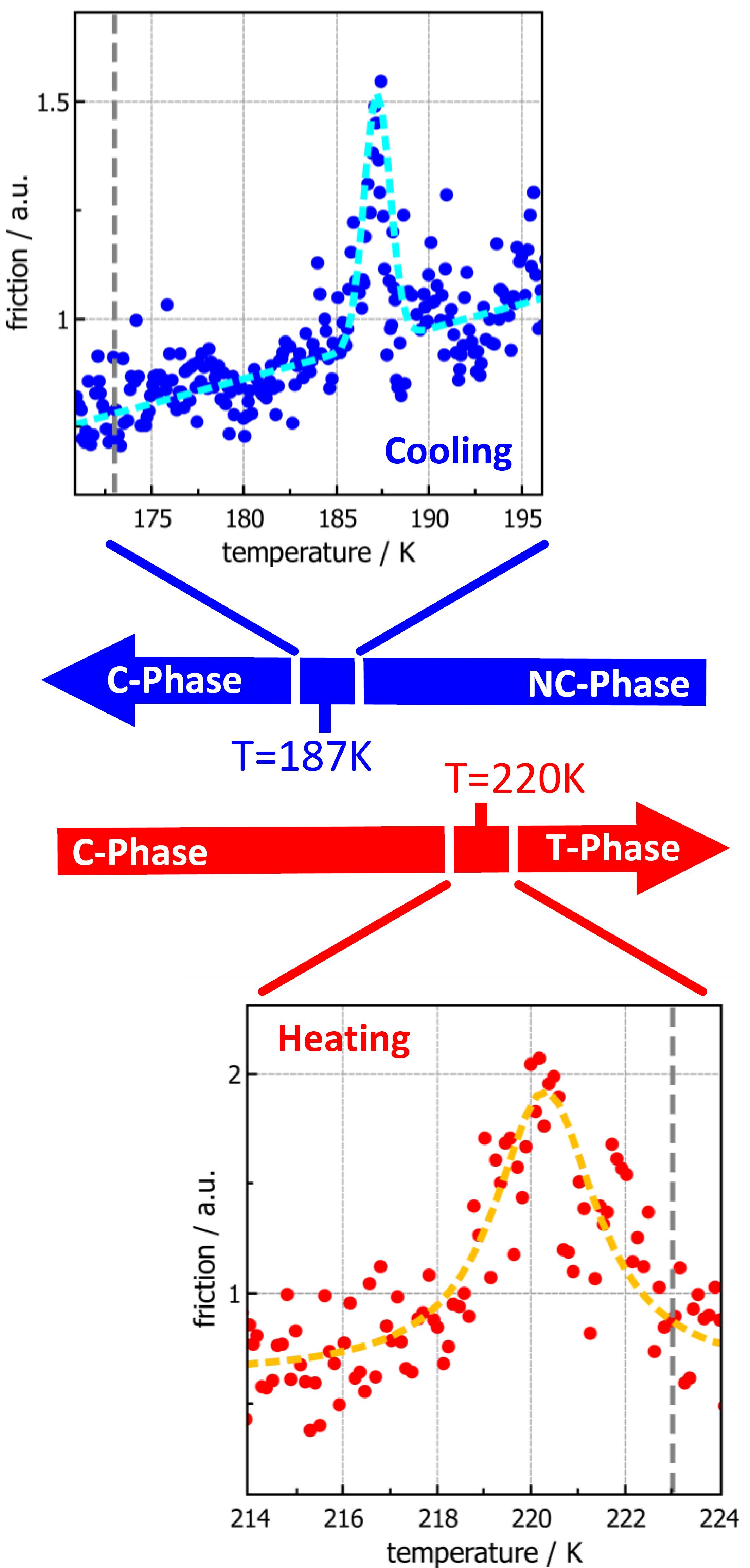}
	\caption{Measured nanoscale friction on 1T-TaS$_2$ as a function of 
sample temperature across the two spinodal transformations. The cooling sequence 
(blue, upper part) shows the NCCDW to CCDW transformation, while the heating 
sequence (red, lower part)  crosses the CCDW to trigonal NCCDW transition. There 
is  no appreciable difference between stable and metastable state friction. 
Between the two, friction shows clear peaks at 186 $\pm 2$K and  at 220 $\pm 
2$K indicating the tip-induced preempting of the bulk spinodal transformations  
at $T_{NC}$ and $T_{CT}$ of 173K and 223K respectively (dashed lines).}  
	\label{fig:heatingcooling}
\end{figure}

For both cooling and heating sequences, the average friction force is calculated 
from each pair of lateral force images recorded for forward and backward 
scanning. Fig. \ref{fig:heatingcooling} shows the resulting friction during 
cooling and heating as a function of the simultaneously recorded temperature. In 
both cases we see a clear peak in the average friction signal at the transition 
temperatures. The peak height is appr 1.5 to 2 times higher than the average 
friction signal away from the transition point, while the peak width is about 2K 
to 5K. Results from further experiments reproduce these values.
In contrast to published contact friction versus temperature results 
\cite{jansen2009nanoscale, jansen2014frictional}, our result shows a very sharp 
and distinct transformation behavior, as is indeed expected from the spinodal 
theory. 

Other details also fall qualitatively in place. The frictional peak on cooling 
occurs near 186 K, which is more than ten degrees higher than $173\pm 2$K, the 
tip-free bulk transformation, assumed to coincide with the spinodal temperature. 
This is precisely what our theory predicts, the temperature difference 
corresponding to $h_c-h_s$, a quantity in principle dependent on details 
including tip size and applied load. Moreover, comparison of heating and cooling 
frictional peaks shows that the heating peak is lower in magnitude and deviates 
less from the bulk temperature $T_{CT} \sim 223$K. This is in agreement with a 
smaller difference expected in this case between $\Psi_m$ and $\Psi_M$, 
reflecting the weaker character known for the transformation on heating  
relative to cooling.\cite{guy1985high} The finite domain size $L$ which limits 
the tip-triggered transformation could in 1T-TaS2 be determined, besides the 
omnipresent defects shown in Fig. \ref{fig:highres}c, also by the recently 
discovered interplanar mosaic structure of this material.\cite{Ma16}

\section {Conclusions}
We have proposed theoretically a mechanism predicting frictional anomalies connected 
with spinodal points which end the hysteresis cycles of first order phase 
transitions. Direct experimental demonstration of  the anomaly is provided  by 
FFM nanofriction measured at the two transformations which occur upon cooling 
(173K) and upon heating (223K) of the NCCDW $\leftrightarrow$ CCDW transition of 
layered 1T-TaS$_2$, transformations which we argue are to a good approximation 
spinodal in character. Near the spinodal temperature the free energy barrier 
protecting the metastable phase decreases enough that the small mechanical 
perturbation provided by the pressing and sliding tip is sufficient to locally 
trigger the  transformation, thus preempting its spontaneous occurrence. The 
frictional anomaly predicted is transient, but can nonetheless be measured in steady 
sliding as the tip explores newer and newer untransformed areas. These  results 
show that nanoscale friction, easy to interpret as it is, is as sensitive as 
resistivity or structural tools such as X-rays, and unlike thermodynamic 
quantities like heat capacity that are totally insensitive when applied to 
spinodal points of first-order phase transitions. In the specific case of 
1T-TaS$_2$, a possible interplay between the known electrical and structural 
characters of the transformations -- characters which apparently do not impact 
the contact friction --  and their spinodal nature, which we exploit here for 
the first time, will deserve  renewed attention in the future. Of special 
interest appears to be for example the possibility to trigger tip-induced 
frictional transformations from hidden states~\cite{Vaskivskyi14}, and/or in the 
ultrathin material, where the spinodal temperature is strongly thickness 
dependent~\cite{Yoshida14,Yoshida15}.

\bibliography{1T-TaS2_friction.bib}
\bibliographystyle{ieeetr}

\begin{acknowledgments}
Work in Trieste was supported through ERC MODPHYSFRICT Contract 320796, also 
benefitted from COST Action MP1303. Financial support in Giessen was provided by 
the German Research Foundation (Project DI917/5-1) and in part by COST Action 
MP1303 and LaMa of JLU Giessen.
\end{acknowledgments}

\end{document}